\documentclass{article}

\begin{document}

\begin{titlepage}

\begin{center}

\vskip 1.5 cm
{\huge \bf
Statistical analysis of the\\[0.2in]
 supersymmetry breaking scale}
\vskip 1.3 cm
{\large Michael R. Douglas$^{1,\&,2}$}\\
\vskip 0.5 cm
{$^1$ NHETC and Department of Physics and Astronomy \\
Rutgers University, \\
Piscataway, NJ 08855-0849, USA\\
$^\&$
I.H.E.S., Le Bois-Marie, Bures-sur-Yvette, 91440 France\\
$^3$
Caltech 452-48, Pasadena, CA 91125, USA}
\vskip 0.3cm
{\tt mrd@physics.rutgers.edu}
\end{center}

\date{June 2004}

\vskip 0.5 cm
\begin{abstract}

We discuss the question of what type and scale of supersymmetry
breaking might be statistically favored among vacua of string/M theory,
building on comments in Denef and Douglas, hep-th/0404116.

\end{abstract}

\thispagestyle{empty}

\end{titlepage}


\vfill\eject

\hyphenation{super-symmetric super-symmetry}
\hyphenation{non-super-symm-etric di-men-sion-al}
\newfam\black
\font\blackboard=msbm10 
\font\blackboards=msbm7
\font\blackboardss=msbm5
\textfont\black=\blackboard
\scriptfont\black=\blackboards
\scriptscriptfont\black=\blackboardss
\def\Bbb#1{{\fam\black\relax#1}}
\newcommand{\eq}[1]{Eq.~(\ref{eq:#1})}
\def\sgn{{\rm sgn\ }}
\def\etal{{\it et.al.}}
\def\slashslash{{/\hskip-0.2em/}}
\def\CA{{\cal A}}
\def\CC{{\cal C}}
\def\CD{{\cal D}}
\def\CF{{\cal F}}
\def\CH{{\cal H}}
\def\CI{{\cal I}}
\def\CJ{{\cal J}}
\def\CL{{\cal L}}
\def\CM{{\cal M}}
\def\CN{{\cal N}}
\def\CO{{\cal O}}
\def\CP{{\cal P}}
\def\CT{{\cal T}}
\def\CW{{\cal W}}
\def\BC{\Bbb{C}}
\def\BH{\Bbb{H}}
\def\BM{\Bbb{M}}
\def\BP{\Bbb{P}}
\def\BR{\Bbb{R}}
\def\BX{\Bbb{X}}
\def\BZ{\Bbb{Z}}
\def\mapr{\mathop{\longrightarrow}\limits}
\def\half{{1\over 2}}
\def\GeV{~{\rm GeV}}
\def\TeV{~{\rm TeV}}
\def\Coh{{\rm Coh}~}
\def\Cohc{{\rm Coh}_c~}
\def\Mod{{\rm Mod}~}
\def\ind{{\rm ind}~}
\def\Tr{{\rm Tr}~}
\def\tr{{\rm tr}~}
\def\grad{{\rm grad}~}
\def\CY#1{CY$_#1$}
\def\rk{{\rm rk}~}
\def\Im{{\rm Im}~}
\def\Hom{{\rm Hom}}
\def\Ext{{\rm Ext}}
\def\Vol{{\rm Vol~}}
\def\Stab{{\rm Stab}}
\def\End{{\rm End~}}
\def\ib{{\bar i}}
\def\jb{{\bar j}}
\def\zb{{\bar z}}
\def\pp{\partial}
\def\pb{{\bar\partial}}
\def\I{{I}}
\def\II{{II}}
\def\IIa{{IIa}}
\def\IIb{{IIb}}
\def\vev#1{{\langle#1\rangle}}
\def\vvev#1{{\langle\langle#1\rangle\rangle}}
\def\ket#1{{|#1\rangle}}
\def\Dslash{\rlap{\hskip0.2em/}D}
\def\dual{{v}} 
\def\intersect{\cdot}
\def\Nbar{{\bar N}}
\def\NRR{{N_{RR}}}
\def\NNS{{N_{NS}}}
\def\bigvev#1{\bigg\langle{#1}\bigg\rangle}
\def\bj{{\bar{j}}}
\def\bi{{\bar{i}}}
\def\bk{{\bar{k}}}

\section{Introduction}

Over the last few years,
a point of view towards the phenomenology of string and M theory has
evolved
\cite{Dine,BouPol,feng,BDM,leshouches,jhs60,KKLT,%
susskind-landscape,mrdstat,ad,BDG,dd,Giryavets:2004zr,DDF,susskind-susy},
which develops and tries to work with the following hypotheses:
\begin{itemize}
\item String/M theory has many consistent vacuum states
which at least roughly match the Standard Model, and
might be candidates to describe our world.
\item The number of vacua is so large that the problems 
of reproducing the Standard Model in detail, and the classic problems
of ``beyond the Standard Model physics'' such as the hierarchy problem
and cosmological constant problem, might admit statistical solutions.
The basic example is that in an ensemble of $N$ vacua which differ
only in a parameter $\Lambda$ (say the c.c. as in \cite{BouPol}),
and in which $\Lambda$ is uniformly
distributed, it is likely that a quantity which appears fine tuned by
an amount $\epsilon > 1/N$ (for the c.c., $10^{-120}$ in a generic
nonsupersymmetric theory) will be realized by at
least one vacuum, just on statistical grounds.
\item No single vacuum is favored by the theory.  Although selection
principles might be found, they will not determine a unique
vacuum {\it a priori}, but rather cut down the possibilities in a way
which is useful only when combined with other information.
\end{itemize}
These three hypotheses are very general and could in principle apply
to any candidate ``theory of everything.''  As such, they
were studied in pre-string work in quantum cosmology.  
The new ingredient is the fourth hypothesis:
\begin{itemize}
\item
String/M theory contains a precisely (and someday even mathematically)
defined set of vacua, and this set has a great deal of structure
which we can to some extent understand.
\end{itemize}
The new idea is that we need to understand this set of possible
vacua, and the larger configuration space containing it, to make
meaningful predictions and test the theory.  Various approaches to
making predictions are under study by various authors,
some combining stringy and anthropic considerations,
others purely statistical.  In any case, one de-emphasizes the study of
individual vacua, in favor of studying distributions of large classes
of vacua.

This structure containing the vacua is often called the stringy
``landscape'' \cite{susskind-landscape},
a term which has long been used in molecular physics,
quantum cosmology, and even mathematical biology (fitness landscapes),
for describing potentials or other functions to be optimized in high
dimensional configuration spaces.  It is a catchy and useful term, so
long as one keeps in mind that the study of the effective
potential is only one aspect of the problem.

Although all of these hypotheses have roots in the early study of
string theory and in quantum cosmology, and in various forms this
point of view has been entertained by many physicists, it seems fair
to say that string theorists have long hoped (indeed most still do)
that this would {\bf not} be the case, that either one of the
remaining problems in constructing a realistic vacuum would prove so
constraining that it alone would rule almost all vacua out of
consideration, or that some fundamental new insight would radically
change the problem and make all considerations based on our present
pictures of string compactification irrelevant.  The main ``conceptual
leap'' required to adopt the new point of view is simply to abandon
these (so far not very productive) beliefs and instead try to work
with the actual results which have come out of string
compactification, which to our mind are quite consistent with the
hypotheses above.

It might even turn out that progress along these lines would reconcile
the old and new points of view.  After all, the problem of finding a
fully satisfactory vacuum is highly constraining; combining the idea
that c.c.  must be tuned, with the many other tests that must be
passed, strongly suggests that the fraction of vacua which work is
significantly less than $10^{-60}$.  We do not know yet that there are
even $10^{60}$ consistent vacua.  What does follow from the
present works is that this ``acceptable'' number of vacua would only
emerge by first making a rough analysis, not enforcing all tests and
consistency conditions, which leads to a embarrassingly large number
of proto-vacua, and then placing many cuts and consistency conditions
on these proto-vacua.

Considering this problem is another way to motivate a statistical
approach, in which one can benefit from the hypothesis than many of
the tests and consistency conditions are to a good approximation
independent, meaning that the fraction of models which pass two tests is
the product of the fractions which pass each test.  While clearly
this is not always true, it is very plausible for many pairs of tests.
For example, while each specific vacuum has a definite number of
generations of quarks and leptons, and a definite cosmological
constant, both of which must agree with observation, there is no
reason to think that these two numbers have any correlation in the
distribution of all vacua.  Extrapolating this type of logic as in
\cite{mrdstat}, one starts to feel it should be far easier to get
useful estimates of the number of compactifications which work, than
to explicitly construct even one complete example.

After setting out some of these ideas in \cite{mrdstat,ad}, I talked to
many phenomenologists to ask what actual statistics of vacua they
might be interested in finding out about.\footnote{
Particular thanks go to S. Dimopoulos, G. Kane, S. Thomas and M. Wise.}
The most common answer was
to learn about the likelihood of low energy supersymmetry and
more specifically the scale of supersymmetry breaking.  

\section{Observations on supersymmetry breaking scales}

Some general considerations were stated in \cite{mrdstat}, and more
specific arguments favoring low scale supersymmetry were given in
\cite{BDG}.

In \cite{dd}, Denef and I studied supersymmetric and non-supersymmetric flux
vacua on CY's with one complex structure modulus in detail, and based
on these results made the following comments in our
conclusions about this problem.

Our considerations were based on the assumption that supersymmetry
breaking is spontaneous and described by the usual $N=1$ supergravity 
Lagrangian, in particular the potential
\begin{equation} \label{eq:defV}
V = e^{{\cal K}/M_p^2}\left( g^{i\bj} D_iW D_{\bj} W^* -
 {3\over M_p^2} |W|^2 \right) + \sum_\alpha D_\alpha^2 .
\end{equation}
Most of our detailed
considerations were for flux vacua in \IIb\ string theory.
\cite{GKP}
However, there are simple intuitive explanations for the claims, so
they could well be more general.

Let us consider a joint distribution of vacua
$$
d\mu[F_i,D_\alpha,\hat\Lambda]
 = \prod dF\ dD\ d\hat\Lambda \rho(F_i,D_\alpha,\hat\Lambda) .
$$
where $\hat\Lambda=3e^K|W|^2$ is the norm of the superpotential,
and $F_i=D_iW$ are the F breaking parameters (auxiliary fields).
The generic vacuum with unbroken supersymmetry would be AdS with
cosmological constant $\Lambda=-\hat\Lambda$.  More generally, 
the c.c. would be
\begin{equation} \label{eq:cc}
\Lambda = \sum_i |F_i|^2  + \sum_\alpha D_\alpha^2 - \hat\Lambda
\end{equation}
and thus the constraint that a physical vacuum should have
$\Lambda\sim 0$ can be expressed as a delta function.  Thus, we can
derive a distribution of supersymmetry breaking scales under the
assumption of near zero cosmological constant, as 
\begin{equation}\label{eq:concc}
d\mu_{\Lambda=0}[F_i,D_\alpha]
 = \prod d^2F_i\ dD_\alpha\ d\hat\Lambda \rho(F_i,D_\alpha,\hat\Lambda) \ %
 \delta(\sum_i |F_i|^2 + \sum_\alpha D_\alpha^2 -\hat \Lambda) .
\end{equation}
Of course, we could solve the delta function for one of the variables.

Now, a main observation of \cite{dd} (section 3.3;
this was suggested earlier by more heuristic
arguments \cite{durham,kachru}) is 
that the distribution of values of $\hat\Lambda$
is uniform near zero, and throughout its range is more or less
uniform.\footnote{
v2: The reason it is $|W|^2$, and not $|W|$ or some other power which
is uniformly distributed, is because (in \cite{dd} and perhaps more generally)
$W$ is uniformly distributed as a complex variable.
Then, writing $W=e^{i\theta}|W|$, we have
$ d^2W = \frac{1}{2} d\theta d(|W|^2)$.}
Furthermore it is relatively uncorrelated with the 
supersymmetry breaking parameters.

The simple physical argument for this is that the superpotential $W$
(and perhaps $K$ as well) gets additive contributions from many
sectors of the theory, supersymmetric and nonsupersymmetric.  If there
are many supersymmetric hidden sectors, varying choices in the hidden
sectors (say of flux) will vary $\hat\Lambda$ in a way which is
uncorrelated with the supersymmetry breaking, and will tend to produce
structureless, uniform distributions.

Thus, we can proceed by solving for $\hat\Lambda$ in \eq{concc}.
Since its distribution is uniform and uncorrelated with $F$ and $D$,
the distribution function $\rho$ is in fact independent of $\hat\Lambda$,
and this simply leads to
\begin{equation}\label{eq:concc2}
d\mu_{\Lambda=0}[F_i,D_\alpha]
 = \prod d^2F_i\ dD_\alpha\ \rho(F_i,D_\alpha) .
\end{equation}

In other words, the constraint of setting the c.c. to a near zero value,
has no effect on the resulting distribution, because we had a wide variety
of supersymmetric effects available to compensate the supersymmetry breaking
contributions.

In itself, this does not prefer a particular scale of supersymmetry
breaking, rather it says that the condition of tuning the c.c. does
{\bf not} lead to a factor of $\Lambda/M_{susy}^4$ favoring low
supersymmetry breaking scales; instead it is neutral with regard to
supersymmetry breaking scale.

We also pointed out that, given many supersymmetry breaking
parameters, the high scale becomes favored.  This simply follows from
the idea that\footnote{
v2: We added D breaking parameters as well; v3: fixed typo.}
\begin{eqnarray}\label{eq:high}\nonumber
d\mu[M_{susy}^2] &= \int \prod_{i=1}^{n_F} d^2F_i\ %
\prod_{\alpha=1}^{n_D} dD_\alpha\ %
d(M_{susy}^4) \delta(M_{susy}^4 - \sum_i |F_i|^2 - \sum_\alpha D_\alpha^2) \\
&\sim (M_{susy}^2)^{2{n_F}+n_D-1} d(M_{susy}^2)
\end{eqnarray}
In words, the total supersymmetry breaking scale is the distance from
the origin in the space of supersymmetry breaking parameters, and
in a high dimensional space most of the volume is near the boundary.
Essentially the same observation is made
independently by Susskind in \cite{susskind-susy}.

This argument is particularly simple if we assume the distributions of
the individual breaking parameters are uniform.  This is more or less
what came out in the one modulus F breaking distributions we studied
in \cite{dd}, but we only considered special limits and this is
probably simplistic.  More generally, it is quite likely that more
complicated regimes or multi-modulus models will contain vacua with
supersymmetry breaking at hierarchically small scales.  

First, the number of vacua in which small scales are generated is
large, although not overwhelmingly so.  In flux compactifications,
this is because of the enhancement of the number of vacua near
conifold points discussed in section 3.1.4 of \cite{dd} (see also
\cite{Giryavets:2004zr}).  While in the one parameter models discussed
in \cite{dd}, these always come with tachyons (in D breaking models),
this phenomenon stemmed from the specific structure of the mass matrix
in one parameter models and is not expected to hold in general.

Second, these effects can drive supersymmetry breaking both by F and D
terms, as seen in many concrete models, for example by putting
antibranes in the conifold throat as discussed in
\cite{KKLT,susskind-susy}.  It seems reasonable to suppose that vacua
with this type of supersymmetry breaking are common, although this
claim remains to be checked.

This should skew the distribution towards small scales but not
overwhelmingly so, as the vast majority of vacua (as seen explicitly
in \cite{dd}) do not produce exponentially small scales.  This is the
sense in which the uniform distribution is still a reasonable picture.

In any case, the underlying observation that adding together many
positive breaking terms as in \eq{high} will produce a distribution
weighed towards high scales is rather general.

\section{Further considerations}

Recently Arkani-Hamed and Dimopoulos have outlined a fairly detailed
phenomenological scenario using high scale supersymmetry breaking
\cite{ahd}.  Rather than the boring prediction that we will only see
the Higgs at LHC, they argue for a very interesting alternate scenario
in which gauginos and Higgsinos could be seen at LHC, mostly on the
grounds that keeping these particles light could duplicate the famous
agreement of supersymmetric grand unification with precision gauge
coupling measurements (see \cite{susygut} for a recent discussion)
in a novel, superficially non-supersymmetric scenario.

While a testable phenomenological scenario is its own justification,
it is also interesting to ask if string theory might favor or disfavor
the new scenario, or better, whether we have any hope of making any
justifiable statement about this question in the near future from
string theory at all.

Now the considerations we just outlined speak against one of the
motivating arguments suggested in \cite{ahd}.  Namely, the idea that
tuning the c.c. is the dominant consideration in analyzing the number
of models.  In the models we discussed, and for the general reasons we
discussed, tuning the c.c. is more or less independent of other
considerations.

On the one hand, we have the standard advantage of low scale supersymmetry
in solving the hierarchy problem.  One would naively guess and
in \cite{ahd} arguments are given that the fraction of vacua which work 
(or tuning factor) is 
\begin{equation}\label{eq:tune}
\rho(M_H,M_{susy}) \propto M_H^2/M_{susy}^2 .
\end{equation}
I believe this is correct.\footnote{
v2: L. Susskind has suggested that even this should not be taken for granted,
because of the well known $\mu$ problem.}

On the other hand, we could have many more high scale models, despite
the neutrality of the c.c. considerations, just because of the many
possible supersymmetry breaking parameters, as expressed in \eq{high}.

What would we need to do to make this precise?  First, we need
to distinguish the question of the overall supersymmetry breaking
scale, from that of the observable supersymmetry breaking scale,
which enters into the solution of the hierarchy problem.

The overall supersymmetry breaking scale, which is essentially
$\hat\Lambda$ defined above, determines the gravitino mass.  It is the
scale which is determined to be high by the previous arguments.
Again, there is a simple intuitive explanation for this claim.  It is
that string compactifications typically contain many hidden sectors,
and there is nothing in our considerations that disfavors
supersymmetry breaking in hidden sectors at a high scale.  Thus one is
led to the general prediction that the gravitino is typically very
heavy, more or less independently of observed physics.

The question of how the many supersymmetry breaking parameters enter
into the observable sector clearly needs a lot of work to understand.
Many mechanisms have been proposed in the literature, and we are now
asking for some distribution over models which realize the various
mechanisms.  This probably requires knowing something about the
distributions of gauge groups and matter content, along the lines
discussed in \cite{mrdstat}.

Since all the basic scenarios discussed in textbooks \cite{kane} are
driven by the expectation value of a single auxiliary field, as a
first guess, one is tempted to say that the parameter which controls
supersymmetry breaking in the observable sector, and thus enters into
\eq{tune}, will be a single $F$ or $D$ term, in other words the
expectation value of the auxiliary component of a single field.

As discussed in the previous section, a reasonable approximation for
the distribution of this term is that it is uniform.  In this case,
vacuum statistics would definitely favor low scale supersymmetry, more
or less by the standard argument.  Indeed, as the standard lore would
have it and as we discussed, the real distribution appears to be
somewhat skewed towards producing exponentially small hierarchies,
which further favors the low scale scenarios.

On the other hand, it could be that in the typical embeddings of the
MSSM and its cousins in string theory, the Higgs mass is determined
by the sum of several positive parameters, analogous to the
sum of squares appearing in \eq{high}, but perhaps fewer in number. 
The resulting distribution would be \eq{high} multiplied by \eq{tune}.
In this case, the high scale would be favored,
even with two $F$ parameters, which would lead to 
$\rho \sim M^4_{susy} d(M_{susy}^2)$.

Taking into account the effects favoring hierarchies and all the
other complexities of the problem clearly requires a more careful
treatment.  An important point which we probably do not understand
well enough to do this properly, is what places the cutoff on the
maximal realized scales of supersymmetry breaking.  One might guess
the Planck scale or the string scale, but the results will probably
be sensitive to this guess.   The specific cutoff which enters 
the flux vacuum counting results in \cite{dd} is the string scale
multiplied by a ``O3 charge'' which is quantized and can reach
$L \sim 1000$ in examples; this is a maximum bound which neglects
many other possible effects.

While this may well be a good estimate for the cutoff on the $W$
distribution (since this receives contributions from supersymmetric
sectors), it seems likely that the cutoff on supersymmetry breaking
scales could be lower.  This is suggested by the
very strong intuition among string theorists that stable
non-supersymmetric models are difficult to construct, which suggests
a cutoff determined by considering stability.

A clear example of this can be seen by considering the masses
of moduli in D-breaking scenarios, as discussed in \cite{DDF}
section 4.  The point made there is the (surely well known) observation
that the bosonic mass matrix in a pure D breaking vacuum is
$$
M^2 = H(H-3e^{K/2}|W|)
$$
where $H=dd(e^{K/2}|W|)$ is the fermion mass matrix.  This is the
generalization of the usual bose-fermi mass relation to a supersymmetric
AdS vacuum, which by assumption is the situation before D breaking, and
the D breaking terms are expected not to change this much.  Thus one
finds that tachyons correspond to fermion masses between $0$ and
$3e^{K/2}|W|=(3\hat\Lambda)^{1/2}$, and the higher the scale of
supersymmetry breaking, the more likely are tachyons.  In itself this
effect probably only lowers the peak of the distribution by a factor
$1/n$ where $n\sim 100$ is the number of moduli, but it is conceivable
that other instabilities are important and drive the scale lower still.

A reasonable summary of the suggestions we have come to is that in
models of supersymmetry breaking which are driven by a single
parameter (F or D term), one should expect low scale breaking, while in
models in which the scale of supersymmetry breaking entering into
\eq{tune} is the sum of squares (or similar combinations) of
more than one independently distributed parameter
(more precisely, $2n_F+n_D\ge 3$), one could expect high scale 
breaking.\footnote{v3: In the borderline case $2n_F+n_D=2$, the
outcome depends on more detailed properties of the distributions.}

\section{Conclusions}

It seems evident that the new approaches to stringy phenomenology
outlined in the introduction, which in the form we are pursuing could
be called the ``statistical approach,'' are stimulating interesting
developments and broadening the range of assumptions used in model
building.

I tried to outline a few of the issues which would go into actually
finding the distribution of supersymmetry breaking scales.  
Many more are discussed in the references.

At this point, it is not at all obvious whether high or low scales
will be preferred in the end.  We explained how this depends on the
coupling of supersymmetry breaking to the observable sector, and using
a simple ansatz of uniform distributions
we stated a model building criterion for cases which could
favor the high scale -- it would be expected if the definition of
``supersymmetry breaking scale'' which controls the Higgs mass (this
need not be the gravitino mass) depends on the sum of independently
distributed positive breaking parameters (one needs the numbers of
F and D parameters to satisfy $2n_F+n_D\ge 3$).

We should keep in mind that ``favoring'' one type of vacuum or
mechanism over another is not a strong result, if both types of vacuum
exist.  It might be that other considerations such as cosmology pick
the less numerous type.  It might be that we just happen to live in
one of the less numerous types.  On the other hand, if the actual
numbers of vacua are not too large, falsifiable claims could come out
of string theory using this approach.  This remains to be seen.

\vskip 0.1in

{\it Note added in v3.} In \cite{newsuss}, Susskind has reviewed these
arguments and suggested that they already predict that supersymmetry
will not be seen at the TeV scale.  This conclusion seems premature as
changing some of the assumptions will reverse the conclusion.  One is
the number of breaking parameters discussed above.  Another is that if
the breaking parameters add with random signs, one does not get the
power law growth of \eq{high}, but instead a narrower distribution
uniform near zero.  Under this assumption, and assuming some models
solve the $\mu$ problem, and combining this with \eq{tune}, in the end
low scale breaking would be favored.

{\it Note added in v4 (following communications with M. Dine, T. Banks,
and S. Shenker)}.
There is a far more serious error in the argument of section 3.
Namely, we considered only the distribution in a sector in which
all supersymmetry breaking parameters are nonzero.  One also expects to
find many partially supersymmetric 
sectors in which only some of the breaking parameters are
nonzero, with the rest zero or very small, 
for reasons explained in \cite{dd} section 4 (since the equations $V'=0$
are quadratic, their solutions lie on several branches, labelled
by the rank of a matrix constructed from the breaking parameters).

While a proper treatment probably requires going into details, a simple
illustration of the possible effects of this is to take the distribution of
breaking parameters suggested in 
section 2 (uniform with a component at hierarchically small scales)
and add a third component which is a delta function at zero.  For purposes
of this argument, the component at hierarchically small scales could also
be taken at zero, leading to 
$d\mu[D] = \delta(D) + c dD|_0^1$
or
$d\mu[F] = \delta(F) + c d(F^2)|_0^1$,
with the parameter $c$ expressing the ratio of the number of supersymmetry
breaking vacua to the sum
of partially supersymmetric vacua and low breaking scale vacua
(in this parameter), and the cutoff taken
to be $1$.  Based on the results
and heuristic arguments in \cite{dd}, one might expect $c\sim 1$.

Convolving these distibutions gives a bimodal distribution with peaks both
at the low and high scale, with (very roughly) 
$d\mu[1]\sim (1+c)^{n_D}(1+2c)^{n_F}$.
This can compensate the tuning factor $M_H^2/M_{susy}^2$, but only with
many breaking terms,  $n \sim |\log_{1+c} M_H^2|$

In words, the simplest reason behind this is that for high scales to
dominate, one needs to start with many more high scale vacua than low
scale or supersymmetric vacua, to compensate for the advantage of low
scale supersymmetry.  Since typically $n>100$ in Calabi-Yau
compactification of string theory, even the revised estimate allows
this to come out, but clearly a believable argument requires more
detail than the simple scaling argument of section 2.  It will be
quite interesting to carry out a discussion of
supersymmetry breaking flux vacua as in \cite{dd} for multiparameter models.

\smallskip

I would like to thank S. Ashok, T. Banks, F. Denef, S. Dimopoulos,
M. Dine, S. Kachru, G. Kane, G. Moore, B. Shiffman, L. Susskind,
S. Thomas, S. Trivedi, L. Wang, M. Wise and S. Zelditch for valuable
discussions.  I particularly thank M. Dine and the entire Stanford
theory group for very lively discussions leading to the v2 comments.

This research was supported in part by DOE grant DE-FG02-96ER40959,
and by the Gordon Moore Distinguished Scholar
program at Caltech.


\begin{thebibliography}{99}

\bibitem{ahd}
N. Arkani-Hamed and S. Dimopoulos, hep-th/0405159.

\bibitem{ad}
S. Ashok and M. R. Douglas, hep-th/0307049.

\bibitem{BDM}
T. Banks, M. Dine and L. Motl,
``On Anthropic Solutions of the Cosmological Constant Problem,''
JHEP 0101 (2001) 031; hep-th/0007206.

\bibitem{BDG}
T. Banks, M. Dine and E. Gorbatov,
hep-th/0309170.

\bibitem{susygut}
W.~de Boer and C.~Sander,
``Global electroweak fits and gauge coupling unification,''
Phys.\ Lett.\ B {\bf 585}, 276 (2004)
hep-ph/0307049.

\bibitem{BouPol}
R. Bousso and J. Polchinski,
JHEP 0006 (2000) 006, hep-th/0004134.

\bibitem{dd}
F.~Denef and M.~R.~Douglas,
``Distributions of flux vacua,''
submitted to JHEP,
hep-th/0404116.

\bibitem{DDF}
F.~Denef, M.~R.~Douglas and B.~Florea,
``Building a better racetrack,''
hep-th/0404257.

\bibitem{Dine}
M. Dine, ``TASI Lectures on M Theory Phenomenology,''
hep-th/0003175.

\bibitem{leshouches}
M. R. Douglas, 
``String Compactification with $N=1$ Supersymmetry,''
in C. Bachas, A. Bilal, M. R. Douglas and N. A. Nekrasov, eds.,
{\it Unity from Duality: Gravity, Gauge Theory and Strings,}
Les Houches 2001, North Holland.

\bibitem{jhs60}
M. R. Douglas, Lecture at JHS60, October 2001, Caltech.
Available on the web at {\tt http://theory.caltech.edu}.

\bibitem{mrdstat}
M.~R.~Douglas,
JHEP {\bf 0305}, 046 (2003)
hep-th/0303194.

\bibitem{durham} M. R. Douglas, Statistics of String Vacua,
to appear in the proceedings of the Workshop on String Phenomenology
at Durham, UK, August 2003,
hep-ph/0401004.

\bibitem{feng}
J. L. Feng, J. March-Russell, S. Sethi and F. Wilczek,
Nucl.Phys. B602 (2001) 307; hep-th/0005276.

\bibitem{GKP}
S. B. Giddings, S. Kachru and J. Polchinski,
Phys. Rev. D66 (2002) 106006; hep-th/0105097.

\bibitem{Giryavets:2004zr}
A.~Giryavets, S.~Kachru and P.~K.~Tripathy,
``On the taxonomy of flux vacua,''
hep-th/0404243.

\bibitem{KKLT}
S.~Kachru, R.~Kallosh, A.~Linde and S.~P.~Trivedi,
hep-th/0301240.

\bibitem{kachru}
S. Kachru, unpublished.

\bibitem{kane}
{\it Perspectives on Supersymmetry},
ed. G. Kane, World Scientific 1998.

\bibitem{susskind-landscape}
L.~Susskind, ``The Anthropic Landscape of String Theory,''
[arXiv:hep-th/0302219].

\bibitem{susskind-susy}
L. Susskind, ``Supersymmetry Breaking in the Anthropic Landscape,''
hep-th/0405189.

\bibitem{newsuss}
L. Susskind, ``Naturalness and the Landscape,''
hep-ph/0406197.

\end{thebibliography}
\end{document}